\documentclass[letter]{aa} 

\usepackage{graphicx}
\usepackage{color}
\usepackage{xcolor}
\usepackage{txfonts}
\newcommand{\kmsM}{km\,s$^{-1}$\,Mpc$^{-1}$}      

\newcommand{\Dt}{\ensuremath{D_{\Delta t}}}                                     \newcommand{\Dd}{\ensuremath{D_\mathrm{d}}}                                     \newcommand{\Ds}{\ensuremath{D_\mathrm{s}}}                                     \newcommand{\Dds}{\ensuremath{D_\mathrm{ds}}}                           
\newcommand{\zd}{\ensuremath{z_\mathrm{d}}}                             \newcommand{\zs}{\ensuremath{z_\mathrm{s}}}  
\newcommand{\Om}{$\Omega_\mathrm{m}$}
\newcommand{\OL}{$\Omega_\Lambda$}
\newcommand{\Ocdm}{$\Omega_\mathrm{CDM}$}
\newcommand{\ODE}{$\Omega_\mathrm{DE}$}
\newcommand{\Ok}{$\Omega_\mathrm{k}$}
\newcommand{\Msun}{M$_\odot$}

\def\hequad{HE\,0435$-$1223}

\def\blens{B1608$+$656}
\def\rxjlens{RXJ1131$-$1231}
\def\sdsslens{SDSS 1206$+$4332}

\begin{document}

   \title{The Hubble constant determined through an inverse distance ladder including quasar time delays and Type~Ia supernovae}

   \author{S.~Taubenberger\inst{1}\thanks{\email{tauben@mpa-garching.mpg.de}}
          \and
          S.\,H.~Suyu\inst{1,2,3}
          \and
          E.~Komatsu\inst{1,4}
          \and
          I.~Jee\inst{1}
          \and
          S.~Birrer\inst{5}
          \and
          V.~Bonvin\inst{6}
          \and
          F.~Courbin\inst{6}
          \and
          C.\,E.~Rusu\inst{7,8,9}
          \and
          A.\,J.~Shajib\inst{5}
          \and
          K.\,C.~Wong\inst{4,7}
          }

   \institute{Max-Planck-Institut f\"ur Astrophysik, Karl-Schwarzschild-Str. 1, 85748 Garching, Germany
         \and
         Physik-Department, Technische Universit\"at M\"unchen, James-Franck-Str. 1, 85748 Garching, Germany
         \and
         Institute of Astronomy and Astrophysics, Academia Sinica, 11F of ASMAB, No.1, Section 4, Roosevelt Rd., Taipei 10617, Taiwan
         \and
         Kavli Institute for the Physics and Mathematics of the Universe (Kavli IPMU, WPI), Univ. of Tokyo, Kashiwa 277-8583, Japan
         \and
         Department of Physics and Astronomy, University of California, Los Angeles, CA 90095-1547, USA
         \and
         Laboratoire d'Astrophysique, Ecole Polytechnique F{\'e}d{\'e}rale de Lausanne (EPFL), Observatoire de Sauverny, 1290 Versoix, CH
         \and
         National Astronomical Observatory of Japan, 2-21-1 Osawa, Mitaka, Tokyo 181-8588, Japan
         \and
         Subaru Telescope, National Astronomical Observatory of Japan, 650 N Aohoku Pl, Hilo, HI 96720
         \and
         Department of Physics, University of California, Davis, 1 Shields Avenue, Davis, CA 95616, USA
         }

   \authorrunning{S. Taubenberger et al.}
   \titlerunning{The Hubble Constant determined though an inverse distance ladder}

   \date{Received ...; accepted ...}

  \abstract
   {The precise determination of the present-day expansion rate of the Universe, expressed through the Hubble constant $H_0$, is one of the most pressing challenges in modern cosmology. Assuming flat $\Lambda$CDM, $H_0$ inference at high redshift using cosmic microwave background data from Planck disagrees at the 4.4$\sigma$ level with measurements based on the local distance ladder made up of parallaxes, Cepheids, and Type Ia supernovae (SNe~Ia), often referred to as Hubble tension. Independent cosmological-model-insensitive ways to infer $H_0$ are of critical importance.}
   {We apply an inverse distance ladder approach, combining strong-lensing time-delay distance measurements with SN~Ia data. By themselves, SNe~Ia are merely good indicators of relative distance, but by anchoring them to strong gravitational lenses we can obtain an $H_0$ measurement that is relatively insensitive to other cosmological parameters.}
   {A cosmological parameter estimate was performed for different cosmological background models, both for strong-lensing data alone and for the combined lensing\,+\,SNe~Ia data sets.}
   {The cosmological-model dependence of strong-lensing $H_0$ measurements is significantly mitigated through the inverse distance ladder. In combination with SN~Ia data, the inferred $H_0$ consistently lies around 73--74\,\kmsM, regardless of the assumed cosmological background model. Our results agree closely with those from the local distance ladder, but there is a $>$2$\sigma$ tension with Planck results, and a $\sim$1.5$\sigma$ discrepancy with results from an inverse distance ladder including Planck, Baryon Acoustic Oscillations, and SNe~Ia.  Future strong-lensing distance measurements will reduce the uncertainties in $H_0$ from our inverse distance ladder.}
   {}

   \keywords{Gravitational lensing: strong --
                cosmological parameters --
                distance scale
               }

   \maketitle
%

\section{Introduction}
\label{Introduction}

Ever since Georges Lema\^itre and Edwin Hubble discovered that our Universe is expanding \citep{lemaitre1927a,lemaitre1931a,hubble1929a}, astronomers have sought to measure the Hubble constant $H_0$ that characterises the present-day expansion rate.  For decades there was the `factor of 2  controversy' in the $H_0$ measurements, culminating in the Hubble Space Telescope Key Project that pinned down $H_0$ to $72\pm8$\,\kmsM\ using the Cepheids distance ladder with several secondary distance indicators including Type Ia Supernovae (SNe~Ia) \citep{freedman2001a,freedman2017a}. In recent years, another controversy on $H_0$ has emerged, particularly between the measurements based on the cosmic microwave background (CMB; $H_0 = 67.36\pm0.54$\,\kmsM\ for flat $\Lambda$CDM; \citealt{planck2018a}) and the local distance ladder (SH0ES programme; $H_0 = 74.03\pm1.42$\,\kmsM; \citealt{riess2019a}). The value of $H_0$ inferred from the CMB depends on the background cosmology, and the 4.4$\sigma$ tension between the Planck and SH0ES measurements refers to a standard flat $\Lambda$CDM cosmology with a spatially flat Universe consisting of cold dark matter and a dark energy that is described by the cosmological constant $\Lambda$.  

This tension, if not resolved by systematic effects \citep[e.g.,][]{rigault2015a,rigault2018a,jones2018a,roman2018a}, indicates new physics beyond flat $\Lambda$CDM \citep[e.g.,][]{divalentino2018a,moertsell2018a,adhikari2019a,agrawal2019a,kreisch2019a,pandey2019a,poulin2019a,vattis2019a}.  Independent measurements of $H_0$ are particularly important in order to  verify this tension, given the potential of discovering new physics.  Methods based on Type~IIP supernova expanding photospheres \citep{schmidt1994b,gall2016a}, water masers \citep{gao2016a, braatz2018a}, or standard sirens \citep{schutz1986a,LIGO2017a} have recently provided independent $H_0$ measurements. While they currently have uncertainties that are consistent with both the Planck and the SH0ES measurements, future measurements with larger samples of Type~IIP supernovae, water masers, and standard sirens could reduce their uncertainties to help shed light on the $H_0$ tension.

Gravitationally lensed quasars are another competitive and independent cosmological probe, particularly in measuring $H_0$.  When a quasar is strongly lensed by a foreground galaxy, multiple time-delayed images of the quasar appear around the lens.  By measuring the time delays between the multiple quasar images and modelling the mass distributions of both the lens galaxy and other structures along the line of sight, strong lensing provides a measurement of the time-delay distance ($\Dt$), which is a combination of the angular diameter distances between the observer and the lens (\Dd), the observer and the quasar (\Ds), and the lens and the quasar (\Dds): $\Dt = (1+\zd) \Dd \Ds / \Dds$ \citep{refsdal1964a, suyu2010a, treu2016a}. While $\Dt$ is inversely proportional and mostly (but not only) sensitive to $H_0$, the inference of $H_0$ from $\Dt$ measurements depends on the cosmological model.  In addition, stellar velocity dispersions of the foreground lens galaxies, if available, provide a determination of $\Dd$, which can further constrain cosmological models \citep[][and in press]{paraficz2009a, jee2015a, jee2016a}.  

The H0LiCOW project  \citep{suyu2017a}, in collaboration with the COSMOGRAIL programme \citep{courbin2018a}, has assembled a sample of lensed quasar systems with exquisitely measured time-delay distances \citep{bonvin2017a, wong2017a, birrer2019a}.  Through a blind analysis, \citet{birrer2019a} reported $H_0=72.5^{+2.1}_{-2.3}$\,\kmsM\ (3\% uncertainty, including systematics) from the data of four lensed quasars, in flat $\Lambda$CDM.  However, as in all cosmological experiments that measure distances outside the scope of the linear Hubble relation $D = cz/H_0$, the inference of $H_0$ from \Dt\ depends on the assumed background cosmology.  While much focus in the community is on $H_0$ in flat $\Lambda$CDM, a cosmological-model-independent inference of $H_0$ is valuable. 

The inverse distance ladder \citep{aubourg2015a, cuesta2015a} provides a way to infer $H_0$ which is more model-independent.  The idea is to anchor the relative distances from SNe~Ia with an absolute distance measurement.  Supernova distances on their own are not absolute distances because of the unknown intrinsic luminosity of SNe \citep[e.g.][]{leibundgut2017a}.  Nonetheless, SNe map out the expansion history of the Universe very precisely, and by anchoring their distance scale with absolute distance measurements, cosmological-model-insensitive absolute distance determinations become feasible.  By anchoring the SN distance scale using distances measured from baryon acoustic oscillations (BAO), \citet{macaulay2019a} measured an $H_0$ from the Dark Energy Survey  consistent with that provided by the \citet{planck2018a} and that does not depend much on cosmological models, although the inference of $H_0$ is strongly affected by the assumptions of the size of the sound horizon \citep{aylor2019a}.  Recently, Jee et al.~(in press) and \citet{wojtak2019a} anchored the SN distance scale using $\Dd$ measured from strongly lensed quasars, resulting in $H_0$ values with $\sim$10$\%$ uncertainty, limited by the precision of the $\Dd$ measurements.  With current data, lensed quasars yield tighter constraints on $\Dt$ than $\Dd$.  In this paper, we explore the use of $\Dt$ as anchor.

This letter is organised as follows.  In Section 2 we summarise the $\Dt$ measurements from the four H0LiCOW lenses, and in Section 3 we use these distances to infer $H_0$ through the inverse distance ladder.  We discuss the results in Section 4, and provide an outlook in Section 5.  Throughout the paper, our parameter constraints correspond to the median values of the parameter probability distributions, with 68\% credibility intervals given by the 16${\rm th}$ and  84${\rm th}$ percentiles.

\section{Lensing time-delay distances}
\label{Lenses}

We use the posterior probability distribution of $\Dt$, $P(\Dt)$, for the four lensed quasar systems that have been measured by the H0LiCOW collaboration (listed in Table \ref{tab:Lenses}).  For three systems (\blens, \rxjlens, and \hequad; \citealt{suyu2010a, suyu2014a}, \citealt{sluse2017a}, \citealt{rusu2017a}, \citealt{wong2017a}, and \citealt{tihhonova2018a}), $P(\Dt)$ is nicely described by the  analytic fit
\begin{equation}
\label{eq:ProbDt}
P(D_{\mathrm \Delta t}) =
\frac{1}{\sqrt{2\pi}(x-\lambda_{\rm D}) \sigma_{\rm D}}
\,\exp \left[ -\frac{(\mathrm{ln}(x-\lambda_{\rm D})-\mu_{\rm D})^{2}}{2\sigma_{\rm D}^{2}} \right],
\end{equation}
where $x=D_{\mathrm \Delta t}/(1\, \mathrm{Mpc})$, and the fitted parameter values ($\lambda_{\rm D}$, $\sigma_{\rm D}$, $\mu_{\rm D}$) are listed in Table 3 of \citet{bonvin2017a}.  For the fourth lens system (\sdsslens), we use the Markov chain Monte Carlo (MCMC) results for $\Dt$ from \citet{birrer2019a}\footnote{The chain of the joint constraint on $\Dt$ and $\Dd$ is released on the H0LiCOW website (http://www.h0licow.org), and we focus on $\Dt$, marginalising over $\Dd$.}, and obtain $P(\Dt)$ through a kernel density estimator.

   \begin{table}
   \caption[]{Lens redshifts ($\zd$) and source redshifts ($\zs$) of the strongly lensed quasars from H0LiCOW included in this study.}
   \label{tab:Lenses}
   \centering
         \begin{tabular}{llll}
            \hline\hline\\[-0.25cm]
            Name            &  $\zd$   &  $\zs$  &  References \\[0.05cm]
            \hline\\[-0.25cm]
            B1608+656       &  0.6304  &  1.394  &  1, 2  \\[0.05cm]
            RXJ1131--1231   &  0.295   &  0.654  &  3, 2  \\[0.05cm]
            HE\,0435--1223  &  0.4546  &  1.693  &  4, 2  \\[0.05cm]
            SDSS\,1206+4332 &  0.745   &  1.789  &  5     \\[0.05cm]
            \hline
         \end{tabular}
         \tablebib{(1)~\citet{suyu2010a}; (2)~\citet{bonvin2017a}; (3)~\citet{suyu2014a}; (4)~\citet{wong2017a}; (5)~\citet{birrer2019a}}
   \end{table}

\section{Inverse distance ladder: supernovae anchored with strongly lensed quasars}
\label{Inverse distance ladder}

To determine the joint likelihood of cosmological parameters for different experiments and cosmological models, we employ the MontePython v3.1 MCMC sampling package \citep{audren2013a,brinckmann2018a}, which is interfaced with the Boltzmann code CLASS \citep{lesgourgues2011a} for CMB calculations. As MCMC algorithm, we select MontePython's Metropolis-Hastings sampler. For every combination of cosmological probes and assumed cosmological background model, we start with a relatively short MCMC chain ($\sim$50\,000 steps) with dynamic updates of the covariance matrix and jumping factor (known as the super-update strategy in MontePython; \citealt{brinckmann2018a}), which ensures an efficient sampling and an acceptance rate close to the optimal 25\%. This is followed by a fully Markovian chain of 500\,000 steps, where the covariance matrix and jumping factor are kept fixed at the previously determined values. Our long chains have acceptance rates between 15\%\ and 30\% and show good convergence.

   \begin{table}
   \caption[]{Priors on cosmological parameters (all uniform) as employed in the MontePython MCMC sampling.}
   \label{tab:Priors}
   \centering
         \begin{tabular}{lcc}
            \hline\hline\\[-0.25cm]
            Parameter    &  Minimum  &  Maximum \\[0.05cm]
            \hline\\[-0.25cm]
            $\Omega_\mathrm{CDM}$  &   0       &  0.45 \\
            $\Omega_\mathrm{k}$    &  $-0.2$   &  0.2  \\
            $w_0$                  &  $-2.5$   &  0.5  \\
            $w_a$                  &  $-2.0$   &  2.0  \\
            $H_0$                  &   0       &  150  \\[0.05cm]
            \hline
         \end{tabular}
   \end{table}

The sampling includes the $H_0$ and \Ocdm\ parameters\footnote{The baryon energy density $\Omega_\mathrm{b}$ is fixed at 0.05, so that \Om\ and \Ocdm\ can be used interchangeably.} and, for cosmological models other than flat $\Lambda$CDM, combinations of \Ok, $w_0$, and $w_a$.  The priors employed for these cosmological parameters are summarised in Table~\ref{tab:Priors}.  They can have an impact on the inferred parameter posteriors, so we make sure that they are either physically motivated or sufficiently conservative.  In those runs where strong-lensing data are combined with SN~Ia data, four additional nuisance parameters ($M_B$, $\alpha$, $\beta$, and $\Delta_M$) are added. They represent the absolute $B$-band magnitude, the coefficients of light curve stretch ($X_1$) and colour ($C$) corrections, and the host-galaxy mass step, respectively, in a SALT2 framework \citep{guy2007a,mosher2014a,betoule2014a}:
\begin{equation}
\label{eq:SALT2}
\mu = m_B - (M_B - \alpha \times X_1 + \beta \times C + \Delta_M).
\end{equation}
The (luminosity) distance modulus, $\mu = 5 \log_{10}(D_{\rm L}/1\,\mathrm{Mpc})+25$, is thereby calculated as the difference between the apparent peak magnitude in the rest frame $B$ band ($m_B$), and the stretch- and colour-corrected absolute $B$-band magnitude. The empirical mass step-correction $\Delta_M$ is only added for SN host galaxies with stellar masses $\geq 10^{10}$\,\Msun\ \citep{sullivan2010a}.

   \begin{table*}
   \caption[]{Cosmological parameters extracted with MontePython MCMC sampling. Results for different cosmological models are shown for quasar time delays alone (upper four lines) and for the combination of the time-delay measurements with the JLA SN~Ia sample (lower six lines). The quoted numbers are the median values, while the uncertainties correspond to the 16$\mathrm{th}$ and 84$\mathrm{th}$ percentiles.}
   \label{tab:Parameters}
   \centering
         \begin{tabular}{lcccccccc}
            \hline\hline\\[-0.25cm]
            Cosmological model    &  \Om  &  \Ocdm  &  \OL  &  \ODE  &  \Ok  &  $w_0$  &  $w_a$ &  $H_0$  \\[0.05cm]
            \hline\\[-0.25cm]
            \textbf{Lenses only:}  \\[0.10cm]
            flat $\Lambda$CDM     & $0.26^{+0.15}_{-0.14}$ & $0.21^{+0.15}_{-0.14}$ & $0.74^{+0.14}_{-0.15}$ & $\equiv0$ & $\equiv0$ & $\equiv-1$ & $\equiv0$ & $72.9^{+2.1}_{-2.3}$  \\[0.15cm]
            flat $w$CDM           & $0.26^{+0.14}_{-0.14}$ & $0.21^{+0.14}_{-0.14}$ & $\equiv0$ & $0.74^{+0.14}_{-0.14}$ & $\equiv0$ & $-1.74^{+0.60}_{-0.45}$ & $\equiv0$ & $80.8^{+5.3}_{-7.1}$  \\[0.15cm]
            flat $w_0w_a$CDM      & $0.27^{+0.14}_{-0.14}$ & $0.22^{+0.14}_{-0.14}$ & $\equiv0$ & $0.73^{+0.14}_{-0.14}$ & $\equiv0$ & $-1.76^{+0.54}_{-0.44}$ & $-0.26^{+1.37}_{-1.21}$ & $81.2^{+5.1}_{-6.3}$  \\[0.15cm]
            non-flat $\Lambda$CDM & $0.26^{+0.15}_{-0.14}$ & $0.21^{+0.15}_{-0.14}$ & $0.72^{+0.16}_{-0.18}$ & $\equiv0$ & $0.03^{+0.12}_{-0.15}$ & $\equiv-1$ & $\equiv0$ & $72.9^{+2.3}_{-2.4}$  \\[0.10cm]
            \hline\\[-0.25cm]
            \textbf{Lenses\,+\,SNe~Ia:}  \\[0.10cm]
            flat $\Lambda$CDM     & $0.30^{+0.04}_{-0.03}$ & $0.25^{+0.04}_{-0.03}$ & $0.70^{+0.03}_{-0.04}$ & $\equiv0$ & $\equiv0$ & $\equiv-1$ & $\equiv0$ & $73.1^{+2.1}_{-2.2}$  \\[0.15cm]
            flat $w$CDM           & $0.28^{+0.10}_{-0.11}$ & $0.23^{+0.10}_{-0.11}$ & $\equiv0$ & $0.72^{+0.11}_{-0.10}$ & $\equiv0$ & $-0.96^{+0.21}_{-0.28}$ & $\equiv0$ & $72.7^{+3.0}_{-2.9}$  \\[0.15cm]
            flat $w_0w_a$CDM      & $0.32^{+0.08}_{-0.11}$ & $0.27^{+0.08}_{-0.11}$ & $\equiv0$ & $0.68^{+0.11}_{-0.08}$ & $\equiv0$ & $-0.97^{+0.20}_{-0.29}$ & $-0.38^{+1.01}_{-1.08}$ & $73.1^{+3.0}_{-3.0}$  \\[0.15cm]
            non-flat $\Lambda$CDM & $0.27^{+0.06}_{-0.05}$ & $0.22^{+0.06}_{-0.05}$ & $0.66^{+0.08}_{-0.06}$ & $\equiv0$ & $0.08^{+0.09}_{-0.13}$ & $\equiv-1$ & $\equiv0$ & $73.4^{+2.2}_{-2.3}$  \\[0.15cm]
            non-flat $w$CDM       & $0.27^{+0.09}_{-0.11}$ & $0.22^{+0.09}_{-0.11}$ & $\equiv0$ & $0.65^{+0.15}_{-0.12}$ & $0.08^{+0.09}_{-0.14}$ & $-1.02^{+0.24}_{-0.34}$ & $\equiv0$ & $73.6^{+3.3}_{-3.3}$  \\[0.15cm]
            non-flat $w_0w_a$CDM  & $0.30^{+0.08}_{-0.10}$ & $0.25^{+0.08}_{-0.10}$ & $\equiv0$ & $0.62^{+0.14}_{-0.11}$ & $0.09^{+0.08}_{-0.14}$ & $-1.05^{+0.25}_{-0.36}$ & $-0.29^{+1.02}_{-1.14}$ & $74.1^{+3.1}_{-3.3}$  \\[0.10cm]
            \hline
         \end{tabular}
   \end{table*}

   \begin{figure*}
   \centering
   \includegraphics[width=17.9cm]{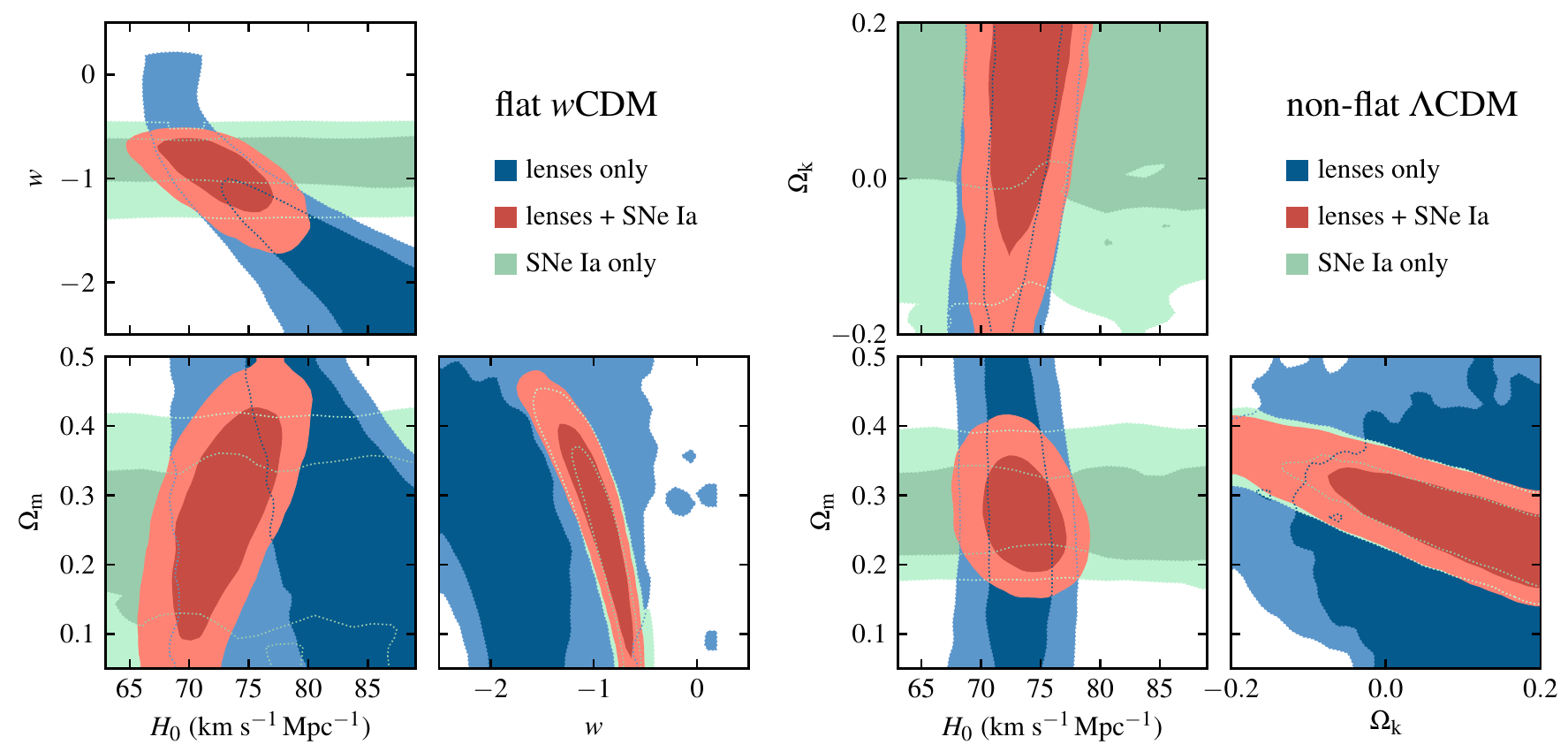}
   \caption{Contour plots with 68\% and 95\% confidence regions for $H_0$, $\Omega_\mathrm{m}$, and $w$ in a flat\,$w$CDM cosmology (left-hand side), and for $H_0$, $\Omega_\mathrm{m}$, and $\Omega_\mathrm{k}$ in a non-flat\,$\Lambda$CDM cosmology (right-hand side). Contours based on quasar time delays and SNe~Ia (JLA compilation) alone are shown in blue and green, respectively, while those using the inverse distance ladder with both probes combined are overplotted in red.}
              \label{fig:corner}%
   \end{figure*}

We first concentrate on the cosmological parameter inference using H0LiCOW \Dt\ data of strongly lensed quasars alone. Four different background cosmologies are investigated: flat $\Lambda$CDM; flat $w$CDM with a redshift-independent dark energy equation-of-state parameter $w$, which can differ from $-1$ (corresponding to $\Lambda$); flat $w_0w_a$CDM with a redshift-dependent dark energy equation-of-state parameter $w(z) = w_0 + w_a \frac{z}{1+z}$; and non-flat $\Lambda$CDM, which covers the possibilities of a negatively or positively curved Universe. The resulting cosmological parameters are shown in Table~\ref{tab:Parameters}. The energy densities of matter (\Om), a cosmological constant (\OL), or a more generic form of dark energy (\ODE) are not tightly constrained by the lensed quasars alone, but the effect of different background cosmologies is very weak for these parameters. For non-$\Lambda$CDM models, $w$ deviates from $-1$ by more than 1$\sigma$, while the curvature in the non-flat $\Lambda$CDM case is consistent with zero. The Hubble constant shows a moderately strong dependence on the background cosmology, in particular on the dark energy   equation of state, being 72.9 in both the flat and non-flat $\Lambda$CDM cases, but $>$\,80 in the flat $w$CDM and $w_0w_a$CDM cosmologies. This can be explained by the lensed-quasar systems spanning redshifts between $\sim$0.3 (for the most nearby lens) and $\sim$1.8 (for the most distant quasar; see Table~\ref{tab:Lenses}), and the necessary extrapolation to obtain the present-day expansion rate of the Universe being cosmological-model-dependent, for example due to degeneracy between $H_0$ and $w$ (Fig.~\ref{fig:corner}).
   
The dependence of $H_0$ on the assumed background cosmology can be mitigated by combining the quasar $\Dt$ measurements with SN data from the joint light curve analysis (JLA), which is a compilation of 740 spectroscopically confirmed low-$z$, SDSS-II, and SNLS SNe~Ia \citep{betoule2014a}. In this inverse distance ladder approach, the SN~Ia data are anchored near their high-$z$ end by the lensed quasars, and allow for a much improved measurement of the present-day expansion rate compared to the quasar time delays alone. We employ the same priors on cosmological parameters as before (Table~\ref{tab:Priors}). In addition to the cosmological models investigated in the lenses-only case, we now also include the more flexible non-flat $w$CDM and non-flat $w_0w_a$CDM models, which did not converge in the lenses-only MCMC chains. The results are again summarised in Table~\ref{tab:Parameters}. The posteriors for \Om\ and \OL\ (or \ODE) have tightened up significantly compared to the lenses-only case, which is a merit of SNe~Ia being able to map out the \textit{\emph{relative}} expansion history of the Universe very well. Similarly, the dark energy equation-of-state parameter $w$ in the non-$\Lambda$CDM models is now better constrained, and very close to $-1$ in all models. The inferred curvature in the non-flat geometries is slightly larger than before, but still consistent with zero. The Hubble constant, finally, shows only a $\sim$2\% variation with the assumed background cosmology, lying between 72.7 and 74.1\,\kmsM\ in all cases. The increased median and uncertainty observed for non-$\Lambda$CDM cosmologies obtained from lensed quasars alone is no longer an issue when combined with SN~Ia data.

   \begin{figure}
   \centering
   \includegraphics[]{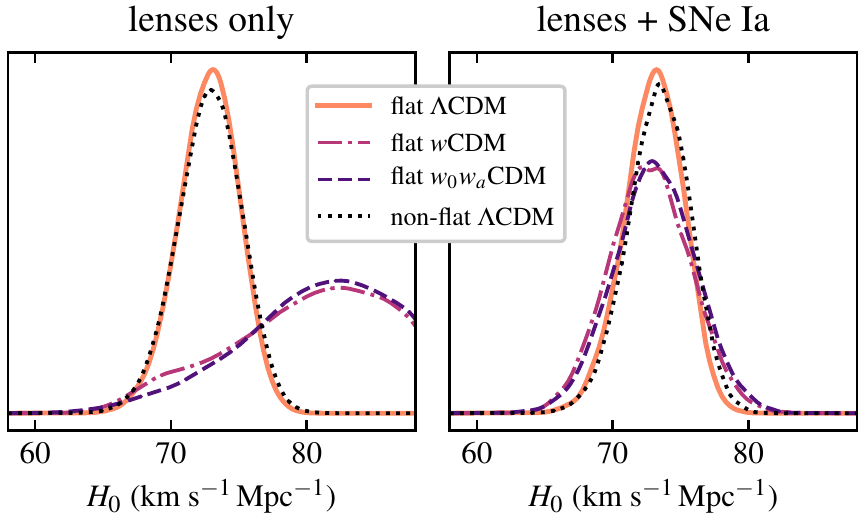}
   \caption{$H_0$ posteriors for different cosmologies using H0LiCOW time-delay distance measurements of four strongly lensed quasars only (left), and using the combination of the lensing measurements with the JLA SN~Ia data set (right).}
              \label{fig:histograms}%
   \end{figure}

For two selected background cosmologies (flat $w$CDM and non-flat $\Lambda$CDM), the full posterior distributions for the sampled cosmological parameters are shown in Fig.~\ref{fig:corner}. The improved constraints on $H_0$ and $w$ (in flat $w$CDM) and on \Om\ (in non-flat $\Lambda$CDM) when including SN~Ia data are evident. Figure~\ref{fig:histograms} compares the marginalised 1D $H_0$ posteriors obtained from lensed quasars alone with those obtained from a combination of lensing and SN Ia data for different cosmological models. Using the inverse distance ladder, the peak-to-peak scatter in the median $H_0$ values of these four models is impressively reduced from 11\% to 1\%.

\section{Discussion}
\label{Discussion}

We now investigate how our $H_0$ results compare to those from other cosmological probes: the SH0ES \citep{riess2019a} and Planck \citep{planck2018a} experiments, and a Planck \,+\,BAO\,+\,SNe~Ia inverse distance ladder \citep{aubourg2015a}.

\begin{figure}
   \centering
   \includegraphics[]{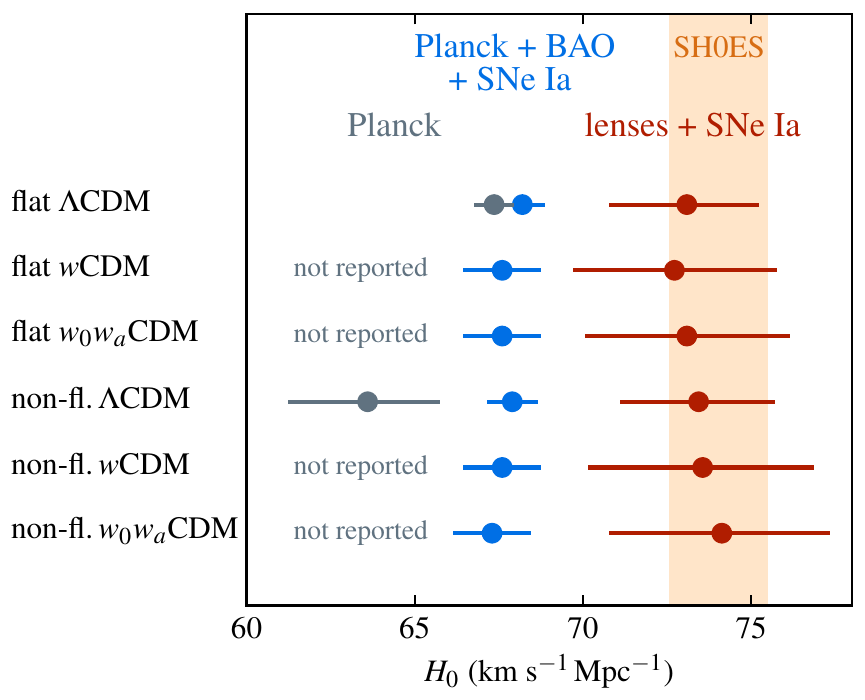}
   \caption{Comparison between the quasar time-delay\,+\,SNe~Ia inverse distance ladder with other cosmological probes: CMB data from Planck (\citealt{planck2018a}; TT,TE,EE\,+\,lowE\,+\,lensing), a Planck\,+\,BAO\,+\,SNe~Ia inverse distance ladder from \citet{aubourg2015a}, and Cepheid\,+\,SN~Ia data from the SH0ES project \citep{riess2019a}. The mean and 68\% confidence intervals for $H_0$ for different background cosmologies are shown for Planck and the two inverse distance ladders. The orange-shaded region reflects the 68\% confidence interval for the SH0ES distance ladder, which is anchored locally and is thus insensitive to the cosmological background model.}
              \label{fig:Hubble tension}%
   \end{figure}

The local distance ladder underlying the SH0ES determination of $H_0$ is anchored to parallaxes at $z\,=\,0$. It is therefore almost completely insensitive to the cosmological background model. As shown in Fig.~\ref{fig:Hubble tension}, our inverse distance ladder measurements of $H_0$ for different cosmologies all agree very nicely with the SH0ES results of $H_0 = 74.03\pm1.42$\,\kmsM. 

Inference of $H_0$ based on CMB, on the contrary, takes place at $z>1000$, and involves extrapolation to $z=0$ by assuming a cosmological model. Hence, while the Planck results for $H_0$ are very precise once a flat $\Lambda$CDM cosmology is assumed, they can vary widely if this assumption is dropped. The \citet{planck2018a} provides $H_0$ values only for flat $\Lambda$CDM ($H_0 = 67.36\pm0.54$\,\kmsM) and non-flat $\Lambda$CDM ($H_0 = 63.6^{+2.1}_{-2.3}$\,\kmsM) cosmologies, and already these differ significantly (see Fig.~\ref{fig:Hubble tension}). With cosmologies that do not assume a cosmological constant, no meaningful constraints on $H_0$ can be obtained from the CMB alone. Our inverse distance ladder results show a tension of just above 2$\sigma$ with Planck for both flat $\Lambda$CDM and non-flat $\Lambda$CDM, which is lower than the tension between Planck and SH0ES owing to our larger error bars compared to SH0ES.

Finally, the Planck\,+\,BAO\,+\,SNe~Ia inverse distance ladder of \citet{aubourg2015a} is anchored to BAO absolute distances at redshifts 0.1--0.6. CMB data are used to set the sound-horizon scale, which BAO distances are inversely proportional to. The inferred values of $H_0$ for different cosmological models are in good agreement with each other, clustering between 67 and 68\,\kmsM, with uncertainties between 1.0\%\ and 1.5\%. The discrepancy with our lensing-based inverse distance ladder is between 1.3 and 1.9$\sigma$, which is not huge, but systematic. A possible origin of this discrepancy could be the adopted sound-horizon scale from Planck, which is the only early-Universe property that enters into the \citet{aubourg2015a} measurement.

\section{Outlook}
\label{Outlook}

On their own SNe~Ia are poor probes of the absolute distance scale of the Universe (and hence $H_0$).  In our inverse distance ladder experiment, where an anchor is provided at high redshift by time-delay distances of strongly lensed quasars, their main role is to extrapolate these absolute distance measurements back to redshift zero.  This allows us to constrain $H_0$ in a way which is 1) rather insensitive to the assumed cosmological background model and 2) independent of Cepheids and the CMB.  The Hubble constant derived from this approach (72.7 to 74.1\,\kmsM) is consistent with that obtained with the local distance ladder, but deviates at the $\sim$1.5--2$\sigma$ level from results based on Planck CMB measurements.  The origin of this discrepancy can only be speculated about, but the sound horizon determined by Planck certainly warrants further investigation.

The small ($\sim$2\%) dependence of $H_0$ on the assumed cosmological model in our inverse distance ladder implies that the \textit{\emph{precision}} of the $H_0$ inference of 3\%--4\% is currently limited by the \Dt\ data for lensed quasars. So far the number of \Dt\ measurements is small: merely four strongly lensed quasars have been published by the H0LiCOW collaboration thus far. However, more are soon to come  (Rusu et al.~submitted, Chen et al.~in prep.), and systematic searches through various surveys\footnote{These surveys include the Dark Energy Survey (DES) \citep[particularly STRIDES;][]{treu2018a}, Gaia, the Hyper-Suprime Cam (HSC) survey, the Kilo-Degree Survey (KiDS), the Panoramic Survey Telescope and Rapid Response System (Pan-STARRS) and the Asteroid Terrestrial-impact Last Alert System (VST-ATLAS)  \citep[e.g.,][]{agnello2018a, krone-martins2018a, lemon2018a, spiniello2018a, rusu2019a}.} are yielding new lensed quasar systems.  Some of these are now being monitored by the COSMOGRAIL collaboration for time-delay measurements \citep[][Millon et al. in prep.]{courbin2018a}.  With the upcoming Large Synoptic Survey Telescope (LSST) and Euclid surveys, many more \Dt\ measurements are  expected, both for strongly lensed quasars and for SNe. Accordingly, the statistical uncertainty on $H_0$ from \Dt\ measurements will shrink substantially in the upcoming years \citep{shajib2018a}, rendering the inverse distance ladder approach that couples time-delay distances with SN~Ia data one of the most promising methods for solving the current Hubble-tension puzzle.

\begin{acknowledgements}
We thank the H0LiCOW team for the public release of \Dt\ likelihoods.  SHS thanks the Max Planck Society for the support through the Max Planck Research Group.  VB and FC acknowledge support from the Swiss National Science Foundation (SNSF).  This project has received funding from the European Research Council (ERC) under the EU’s Horizon 2020 research and innovation programme (grant agreements No 771776 and No 787866). We thank Thejs Brinckmann for the helpful hints on how to run MontePython, Andreas Weiss for computing support, and the anonymous referee for constructive comments.
\end{acknowledgements}

\bibliographystyle{aa}

\end{document}